\documentclass[lettersize,journal]{IEEEtran}
\usepackage{amsmath,amsfonts}
\usepackage{array}
\usepackage[caption=false,font=normalsize,labelfont=sf,textfont=sf]{subfig}
\usepackage{textcomp}
\usepackage{stfloats}
\usepackage{url}
\usepackage{verbatim}
\usepackage{graphicx}
\usepackage{cite}
\usepackage{lineno}
\usepackage{float}
\usepackage{multirow}
\usepackage{caption}
\usepackage{adjustbox}
\usepackage{tabularx}

\usepackage{algorithm} 
\usepackage{algpseudocode}
\usepackage{color}

\usepackage{hyperref}
\begin{document}

\title{Real-Time Forecasting of Dockless Scooter-Sharing Demand: A Spatio-Temporal Multi-Graph Transformer Approach}

\author{Yiming Xu, Xilei Zhao, Xiaojian Zhang, Mudit Paliwal
\thanks{Corresponding author: Xilei Zhao (xilei.zhao@essie.ufl.edu)}
\thanks{This work has been submitted to the IEEE for possible publication. Copyright may be transferred without notice, after which this version may no longer be accessible.}
        }

\markboth{IEEE Transactions on Intelligent Transportation Systems}%
{Shell \MakeLowercase{\textit{et al.}}: A Sample Article Using IEEEtran.cls for IEEE Journals}

\IEEEpubid{}

\maketitle

\begin{abstract}
Accurately forecasting the real-time travel demand for dockless scooter-sharing is crucial for the planning and operations of transportation systems. Deep learning models provide researchers with powerful tools to achieve this task, but research in this area is still lacking. This paper thus proposes a novel deep learning architecture named Spatio-Temporal Multi-Graph Transformer (STMGT) to forecast the real-time spatiotemporal dockless scooter-sharing demand. The proposed model uses a graph convolutional network (GCN) based on adjacency graph, functional similarity graph, demographic similarity graph, and transportation supply similarity graph to attach spatial dependency to temporal input (i.e., historical demand). The output of GCN is subsequently processed with weather condition information by the Transformer to capture temporal dependency. Then, a convolutional layer is used to generate the final prediction. The proposed model is evaluated for two real-world case studies in Washington, D.C. and Austin, TX, respectively, and the results show that for both case studies, STMGT significantly outperforms all the selected benchmark models, and the most important model component is the weather information. The proposed model can help the micromobility operators develop optimal vehicle rebalancing schemes and guide cities to better manage dockless scooter-sharing operations.
\end{abstract}


\section{Introduction}
\IEEEPARstart{S}{hared} micromobility refers to small, single-passenger transportation modes rented for short-term use, such as dockless bike-sharing and scooter-sharing. Shared micromobility is flexible, convenient, affordable, environmentally friendly, and fun to use, making it especially attractive to serve short-distance trips and offer a potential solution to the “first mile/last mile” problem that has long troubled public transit. Among all the shared micromobility options, dockless scooter-sharing is growing at the fastest pace \cite{nacto2019a}. In 2019, people in the U.S. took 86 million trips on dockless scooter-sharing systems, contributing to an over 100\% increase from 2018 \cite{nacto2019a}. 
While dockless scooter-sharing as a travel mode can greatly enhance urban mobility, it is faced with two major operational challenges. First, the trip origin and destination demands of dockless scooter-sharing are spatially and temporally unbalanced \cite{mckenzie2019spatiotemporal, merlin2021segment, xu2020micromobility}. Since urban areas have limited parking spaces for scooters, unbalanced trip demand would lead to the gathering of scooters in certain places and spilling out of the parking space, thus blocking sidewalks and causing safety issues. The gathering of scooters should be anticipated so that the redundant scooters are removed in time to mitigate the adverse effects. In light of this, accurate spatiotemporal dockless scooter-sharing demand predictions are essential for generating optimal vehicle rebalancing strategies \cite{schuijbroek2017inventory} and removing the redundant scooters. Second, the scooters need to be recharged when they are at low power level. To accomplish this, the operators collect these low-power-level vehicles, recharge them, and then drop them off at specific locations \cite{moreau2020dockless}. In this case, accurate dockless scooter-sharing demand predictions are essential to determine the optimal scooter drop-off locations.

Although some recent studies (e.g., \cite{mckenzie2019spatiotemporal, bai2020dockless, zhu2020understanding}) have examined the spatiotemporal usage patterns of dockless scooter-sharing in different cities, few studies shed light on the highly-accurate real-time spatiotemporal dockless scooter-sharing demand prediction. Additionally, most real-time shared micromobility demand forecasting studies are focused on station-based (i.e., docked) bike-sharing \cite{chen2020predicting, lin2018predicting, li2019learning, pan2019predicting}. Based on models such as recurrent neural network (RNN), convolutional neural network (CNN), and graph convolutional network (GCN), these studies abstracted the stations into nodes in a graph and assigned the demand to these nodes. Although these models achieved a decent prediction performance for station-based bike-sharing demand forecasting, applying these models to dockless scooter-sharing demand forecasting can be challenging since there are no parking stations for dockless scooter-sharing. The station-based bike-sharing trips can only take place at the stations, while dockless scooter-sharing trips can happen at any place where no scooter parking restrictions exist. The origin and destination demand of station-based bike-sharing is concentrated in the stations, while the spatial distribution of dockless scooter-sharing demand is more scattered. There is a pressing need to study spatiotemporal demand forecasting for dockless scooter-sharing services.
 
In this paper, we propose a graph-based deep learning model named \textit{Spatio-Temporal Multi-Graph Transformer} (STMGT) to deal with the real-time dockless scooter-sharing demand forecasting problem. The proposed model simultaneously captures spatial and temporal dependency to achieve high prediction performance. Four graphs representing the spatial correlations (i.e., spatial adjacency, functional similarity, demographic similarity, and transportation supply similarity) are first constructed. Then we use a GCN module based on the four graphs to attach the spatial dependency to the input time series data (i.e., historical demand). After that, a sequence of Transformer blocks is used to process the output of the GCN module and the weather condition information to capture temporal dependency. At last, a $1\times 1$ convolutional layer is used to generate the final prediction. The proposed architecture which utilizes Transformer and multi-graph GCN significantly outperforms the state-of-the-art benchmark models for two separate case studies (Washington, D.C. and Austin, TX). Particularly, the input graphs are specifically designed for the dockless scooter-sharing demand forecasting problem using transportation domain knowledge. The major contributions of this study are two-fold: 1) we propose a novel deep learning framework that effectively integrates Transformer and multi-graph GCN to deal with an important but rarely studied problem (i.e., real-time dockless scooter-sharing demand forecasting); and 2) we incorporate transportation domain knowledge (i.e., including a number of new features associated with dockless scooter-sharing demand) in the deep learning model to achieve a high prediction accuracy. 


The remainder of this paper is structured as follows. Section~\ref{s:2} reviews existing literature related to this study. Section~\ref{s:3} formally defines the research problem and describes the proposed method from the overall model architecture to its specific components. Section~\ref{s:4} presents case studies in Washington, D.C. and Austin, TX, and compares the performance of the proposed model with several benchmark models. Section~\ref{s:5} concludes the paper by summarizing findings, identifying limitations, and suggesting future work.

\section{Literature Review}
\label{s:2}
\subsection{Deep-Learning-Based Travel Demand Forecasting Models}
The travel demand forecasting problem has been intensively studied by researchers in the past several decades. A number of forecasting models have been proposed to tackle this problem. These models can be roughly categorized into three categories: the statistical models, the classical machine learning models, and the deep learning models. In this section, we focus on the deep learning models.



The deep learning methods have provided researchers powerful tools to deal with travel demand prediction problems, such as taxi demand prediction \cite{liu2019contextualized, xu2017real, yao2018deep}, ride-hailing demand prediction \cite{geng2019spatiotemporal}, ridesourcing demand prediction \cite{ke2021predicting}, and bike-sharing demand prediction \cite{lin2018predicting, li2019learning,kim2019graph}. Since the demand varies spatially and temporally, different deep learning methods were used to capture spatial dependency and temporal dependency in these studies, and the results showed that these deep learning methods outperformed the classical machine learning models (e.g., random forest and gradient boosting decision tree) and the statistical models (e.g., linear regression and ARIMA). In these demand prediction tasks, the convolutional neural networks (CNNs) and graph convolutional networks (GCNs) are usually used to capture the spatial dependency, and the recurrent neural networks (RNNs) such as gated recurrent unit (GRU) and long short-term memory (LSTM) models are usually used to capture the temporal dependency \cite{chen2020predicting, pan2019predicting}. Some studies also combined these two kinds of neural networks into a spatiotemporal model to capture spatial and temporal dependency simultaneously, and the results showed that the spatiotemporal models have better performance than the single CNN or RNN models \cite{tang2021multi, geng2019spatiotemporal, yao2018deep, yu2018spatio, bai2020adaptive}. However, these convolution-based recurrent models suffer from time-consuming training process and limited scalability for modeling long sequences \cite{cai2020traffic, xu2020spatial}. Based on attention mechanism and encoder-decoder structure, attention-based models such as Transformer \cite{vaswani2017attention} and Informer \cite{zhou2021informer} can efficiently capture long-range temporal dependency from sequential data. Researchers have combined Transformer with spatial models (e.g., GCNs) to solve traffic flow forecasting problems \cite{xu2020spatial, cai2020traffic} and ride-hailing demand forecasting problems \cite{li2021intercity}. However, studies on applying Transformer to micromobility demand forecasting problems are lacking. 

Given the effectiveness of deep learning methods on demand forecasting tasks, different kinds of deep learning models have been applied for shared micromobility demand prediction in previous studies, including the RNN models \cite{chen2020predicting, pan2019predicting} and the spatiotemporal models \cite{lin2018predicting, liu2019contextualized}. Although some of these models can capture spatial and temporal dependency, they only consider the geographical adjacency relationships between zones when extracting spatial dependency. Many spatial factors, such as the zonal functionality and demographic and built environment characteristics (e.g., population density) that are highly related to passenger trip demand \cite{bai2020dockless, xu2021identifying} are omitted in these models. In addition, the usage of shared micromobility is sensitive to weather conditions \cite{noland2021scootin, younes2020comparing}, but few existing spatiotemporal models took the weather conditions into account \cite{kim2019graph}. To fill these gaps mentioned above, we develop a spatiotemporal model based on Transformer and GCN. The model takes the weather conditions into account and constructs several graphs to include more spatial information (e.g., demographic characteristics) in the model to achieve high prediction accuracy.

\subsection{Factors Associated with Dockless Scooter-Sharing Usage}

Many factors have been examined by the researchers to understand the influence of these factors on dockless scooter-sharing usage. These factors can be categorized as weather conditions, demographic factors, built environment factors, and land use factors. 

Literature showed that the weather conditions, such as temperature, precipitation, and wind speed, can greatly influence the usage of dockless scooters \cite{noland2021scootin, younes2020comparing, mathew2019impact}. For example, the rainy and cold weather will significantly reduce the use of dockless scooter-sharing services \cite{noland2021scootin}. 

The demographic factors include age, gender, income, education level, race, resident status, and so on. The literature suggested that the young people, the male, the people with high income, and highly educated people were more likely to use the dockless scooter-sharing services \cite{lee2021factors, cao2021scooter, mitra2021potential, christoforou2021using, laa2020survey, sanders2020scoot}. Dockless scooter-sharing usage was also found to be positively associated with some zonal demographic characteristics, including population density, employment rate, proportion of young population, and proportion of highly educated population \cite{merlin2021segment, bai2020dockless, caspi2020spatial}. 

The most significant built environment factor related to the usage of dockless scooter-sharing was the transportation supply factors, such as the quality of riding environment, especially the street safety \cite{mitra2021potential, sanders2020scoot, hosseinzadeh2021scooters}, and access to transit stations. Areas with better riding environment (i.e., higher Walk Score and Bike Score, and better bicycle infrastructure) often had a high density of shared-scooter trips \cite{hosseinzadeh2021scooters, caspi2020spatial}. Areas with higher transit station density usually had more shared-scooter trips \cite{bai2020dockless, merlin2021segment}. 

Land use factors were also associated with the spatial usage patterns of shared-scooters: greater land use diversity and higher proportion of commercial land use were positively correlated with shared-scooter demand \cite{hosseinzadeh2021scooters, bai2020dockless, merlin2021segment}. 

Since the factors discussed above can greatly influence the usage of dockless scooter-sharing, these factors should be considered when forecasting the dockless scooter-sharing demand. Although some existing demand forecasting models have taken some of these factors into account \cite{tang2021multi,ke2021predicting}, few studies comprehensively included all kinds of these factors in their models when predicting the real-time dockless scooter-sharing demand. 

\section{Methodology}
\label{s:3}
\subsection{Research problem definition}
In this section, we formally define the zonal dockless scooter-sharing demand prediction problem and introduce several key inputs (i.e., dockless scooter-sharing demand, node correlation graph, historical demand matrix, and weather condition matrix) used in our study.

\textbf{\textit{Definition 1}}: Dockless scooter-sharing demand $x_t^i$. We first divide the study area into several zones. In previous studies on demand prediction \cite{geng2019spatiotemporal, yao2018deep, liu2019contextualized}, the study area was usually divided into regular cells (e.g., squares and hexagons). Although that kind of segmentation enables the use of standard machine learning algorithms (e.g., CNNs), it cannot well represent the functional and administrative properties of the zones \cite{ke2021predicting}. Therefore, in this study, we divide the study area by the census block groups and census tracts, in which the socioeconomic and demographic properties are homogeneous. After that, we count the number of dockless scooter-sharing uses in each area during a specific time interval. The count is defined as dockless scooter-sharing demand. The dockless scooter-sharing demand of an area $i$ at time interval $t$ is denoted by $x_t^i$. 

\textbf{\textit{Definition 2}}: Node correlation graph $G$. We use an unweighted graph $G=(V,E)$ to describe the spatial and property correlation between the nodes (i.e., census block groups or census tracts). This graph is fused by multiple graphs to represent the spatial adjacency relationship, functional similarity, demographic similarity, and transportation supply similarity between the zones. In graph $G=(V,E)$, $V=\{v_1,v_2,...,v_N\}$ is a set of nodes (i.e., census block groups or census tracts), where $N$ is the number of nodes. $E$ is a set of edges. If two nodes in $G$ are correlated (e.g., spatially adjacent or have similar functionality), there is an edge between these two nodes. An adjacency matrix $A \in \mathbb{R}^{N \times N}$ can be used to represent the graph $G$. The element $a_{ij} \in A$ is 1 if there is an edge between node $i$ and node $j$; otherwise, 0. The details of node correlation modeling are discussed in Section~\ref{node_corr}.

\textbf{\textit{Definition 3}}: Historical demand matrix $\textbf{X}^{N \times T}$. In historical demand matrix $\textbf{X}^{N \times T}$, $N$ is the number of nodes (i.e., census block groups or census tracts), and $T$ is the sequence length of input historical demand data. $\textbf{X}_t = [X_{t-T+1},\dots,X_t]$ where $X_t=[x_t^1,x_t^2,...,x_t^N]$ denote the dockless scooter-sharing demand of all nodes at time $t$.

\textbf{\textit{Definition 4}}: Weather condition matrix $C^{N \times M_w}$. The dockless scooter-sharing demand can be greatly influenced by the weather. We use a weather matrix $C^{N \times M_w}$ to represent the weather conditions of nodes, where $N$ is the number of nodes and $M_w$ is the number of weather features (e.g., temperature, precipitation, and wind speed).\\

Based on the definitions above, the zonal dockless scooter-sharing demand prediction problem is formulated as follows:

\textbf{\textit{Problem}}: Given a node correlation graph $G$, a weather condition matrix $C$, and a historical demand matrix $\textbf{X}_t$, learn a function $f:\mathbb{R}^{N \times T} \rightarrow \mathbb{R}^{N\times M}$ that maps historical dockless scooter-sharing demand of all zones to the demand in next $M$ time intervals:
\begin{equation}
\textbf{Y}_t = [X_{t+1},\dots,X_{t+M}] = f(\textbf{X}_t,G,C)
\end{equation}

\subsection{Overview of model framework}

\begin{figure*}[hbt!]
    \centering
    \includegraphics[width=0.9\textwidth]{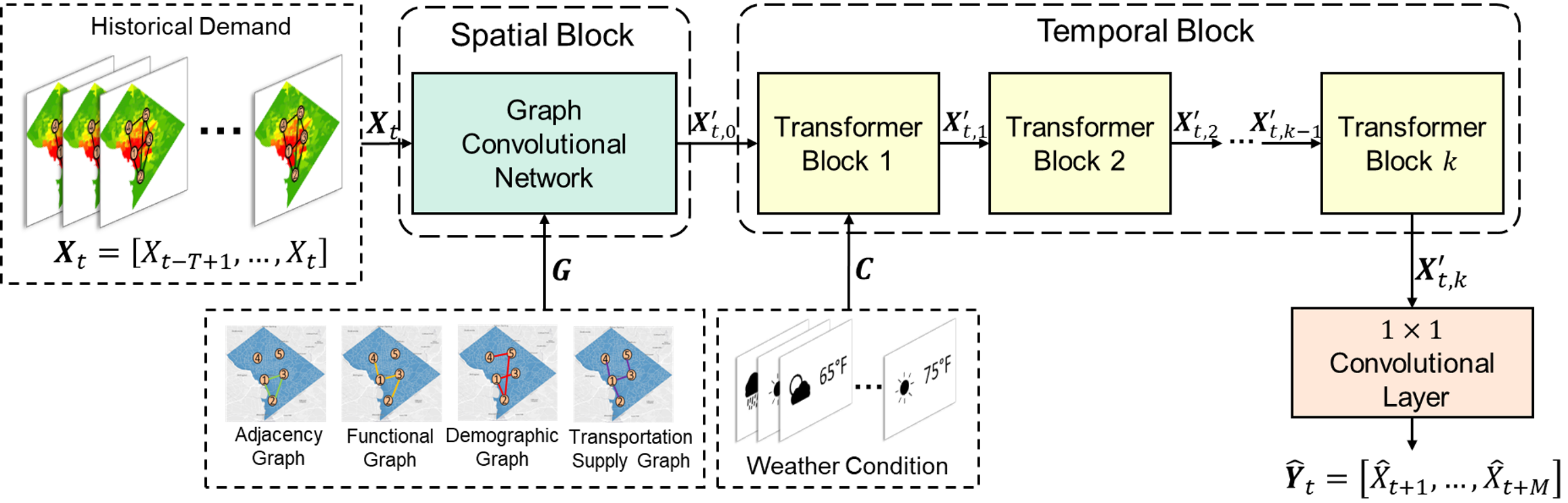}
    \caption{Overall Model Framework}
    \label{fig:frwk}
\end{figure*}

We propose a \textit{Spatio-Temporal Multi-Graph Transformer} (STMGT) model to solve the research problem. The model framework is presented in Figure~\ref{fig:frwk}. The model is composed of a spatial block, a temporal block, and a $1 \times 1$ convolutional layer. The spatial block is a graph convolutional network (GCN) based on node correlation graph $G$. The node correlation graph $G$ is generated by four graphs including adjacency graph, functional similarity graph, demographic similarity graph, and transportation supply similarity graph. The spatial block takes historical demand $\textbf{X}_t = [X_{t-T+1},\dots,X_t]$ as input, and generates block output $\textbf{X}_{t,0}'$ by attaching spatial information from node correlation graph $G$ to $\textbf{X}_t$. The temporal block is composed of a sequence of $k$ Transformer blocks. The first Transformer block takes weather condition matrix $C$ and the output of spatial block $\textbf{X}_{t,0}'$. The subsequent Transformer block takes the output of the previous transformer block. The output of temporal block $\textbf{X}_{t,k}'$ is subsequently processed by a $1 \times 1$ convolutional layer to generate the prediction $\widehat{\textbf{Y}}_t = [\widehat{X}_{t+1},\dots,\widehat{X}_{t+M}]$. 

\subsection{Graph Convolutional Network (GCN)}
\label{sec:gcn}
Extracting correlation dependency between different zones can enhance the performance of the dockless scooter-sharing demand prediction. For example, the neighborhood zones are more likely to have similar dockless scooter-sharing usage patterns, and that information can be used in demand prediction to improve the performance of the model. In previous studies, both convolutional neural network (CNN) and graph convolutional network (GCN) can be used to capture the spatial dependency \cite{zhao2019t,yao2018deep,geng2019spatiotemporal}. However, the CNN can only be performed in Euclidean space, such as images and square grids, while GCN can handle graph-structured data \cite{zhao2019t}. Since the census block groups do not have a regular spatial structure (with varying shapes and areas), they can be easily represented by a graph, and thus GCN is chosen in this study. 

Given an adjacency matrix $A$ and the historical demand matrix $\textbf{X}_t$, GCN performs convolutional operation using a filter in the Fourier domain. The filter is applied to each node of the graph, thus capturing spatial dependency between this node and its adjacent nodes. The GCN model can be constructed by stacking multiple convolutional layers:
\begin{equation}
H^{l+1}=\sigma (\tilde{D}^{-\frac{1}{2}} \tilde{A} \tilde{D}^{-\frac{1}{2}} H^l W^l)
\end{equation}
\noindent where $H^l$ is the output of layer $l$ and $H^0=\textbf{X}_t$, $\tilde{A}=A+I$ is the adjacency matrix of the graph $G$ with self-connections, $I$ is the identity matrix, $\tilde{D}$ is the diagonal node degree matrix of $\tilde{A}$, and $W^l$ is a layer-specific trainable matrix. $\sigma(\cdot)$ denotes an activation function, such as the $ReLU(\cdot)=max(0,\cdot)$ \cite{nair2010rectified}. In this study, we use a 2-layer GCN model introduced by Kipf and Welling (2016) \cite{kipf2016semi} to capture node correlation dependency. We first calculate $\widehat{A}=\tilde{D}^{-\frac{1}{2}} \tilde{A} \tilde{D}^{-\frac{1}{2}}$ in a pre-processing step. The forward model then takes the form:
\begin{equation} \label{eq:z}
\textbf{X}_{t,0}'=f(\textbf{X}_t,A)=softmax(\widehat{A}~ReLU(\widehat{A}\textbf{X}_tW^0)~W^1)
\end{equation}
\noindent where $W^0 \in \mathbb{R}^{C \times H}$ is the input-to-hidden weight matrix, $C$ is the number of input channels (i.e., a $C$-dimensional feature vector for each node), $H$ is the number of hidden units, $W^1 \in \mathbb{R}^{H \times F}$ is the hidden-to-output weight matrix, $F$ is the number of filters, $\textbf{X}_{t,0}' \in \mathbb{R}^{N \times F}$ is the output convolved matrix, and $N$ is the number of nodes. The softmax activation function, defined as $softmax(x_i)=exp(x_i)/\sum_i exp(x_i)$, is applied row-wise.

\subsection{Transformer}
\label{sec:tranformer}
Transformer is a deep learning network based on attention mechanisms \cite{vaswani2017attention}. With the parallelizable self-attention mechanism, Transformer can adaptively capture long-range temporal dependencies from sequential data, thus shows great sequence learning ability in various time series modeling applications \cite{vaswani2017attention, xu2020spatial, cai2020traffic, zhou2021informer, li2019enhancing}. 

\begin{figure}[hbt!]
    \centering
    \includegraphics[width=0.37\textwidth]{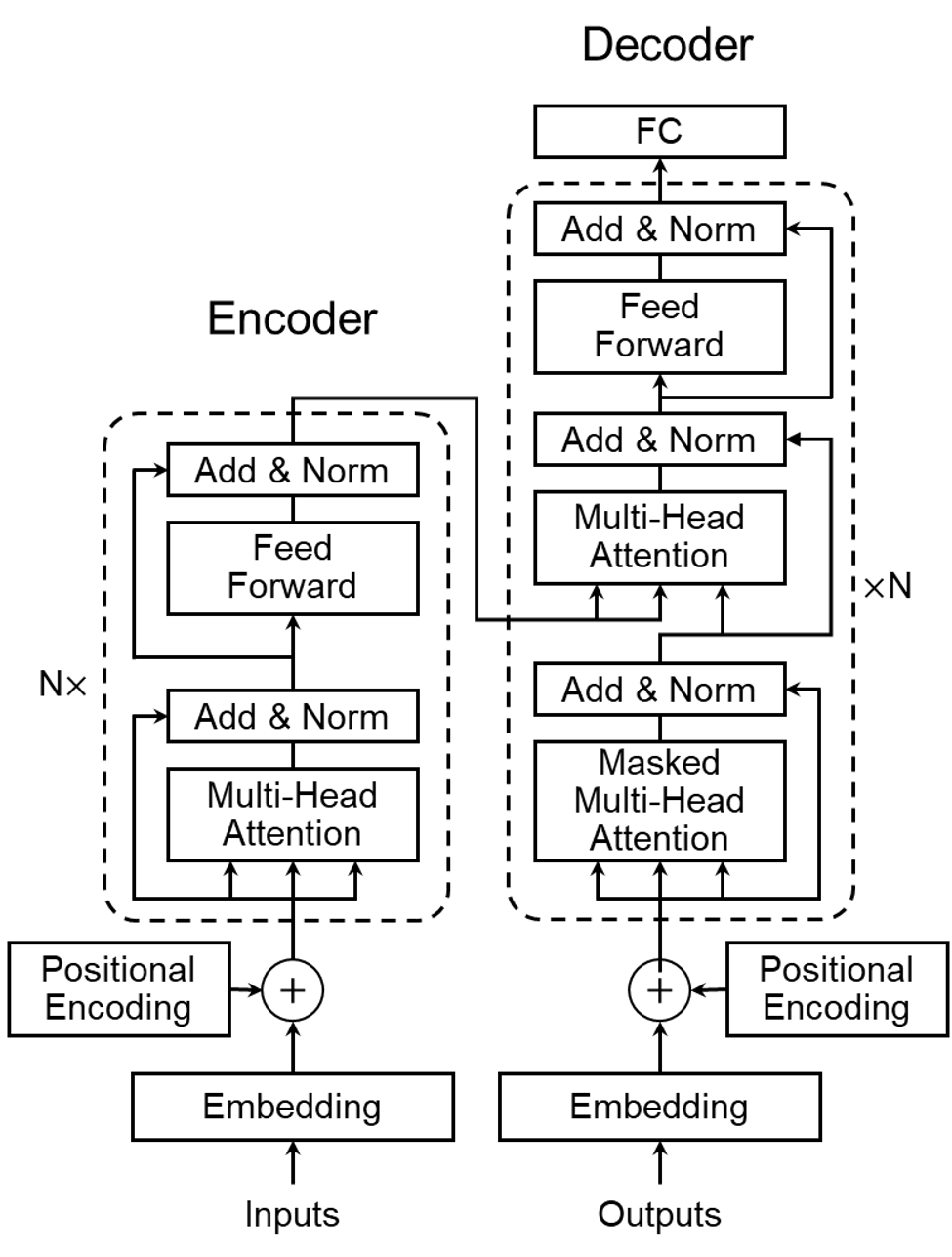}
    \caption{The Transformer Architecture (adapted from \cite{vaswani2017attention})}
    \label{fig:tran_frwk}
\end{figure}

The overall architecture of the transformer is presented in Figure~\ref{fig:tran_frwk}. The transformer is composed of an encoder and a decoder. The encoder is a stack of $N$ identical layers. Each layer has two sub-layers. The first sub-layer is a multi-head self-attention mechanism, and the second is a position-wise fully connected feed-forward network. A residual connection \cite{he2016deep} is employed around both sub-layers, followed by layer normalization \cite{ba2016layer}. The encoder outputs a vector representation for each position of the input sequence. The decoder is also a stack of $N$ identical layers with residual connections and layer normalizations. In addition to the two sub-layers in the encoder, the decoder inserts a third sub-layer, which performs multi-head attention over the output of the encoder stack. The self-attention sub-layer in the decoder stack is also masked to prevent positions from attending to subsequent positions. 

The attention mechanism maps a query and a set of key-value pairs to an output. The output is computed as a weighted sum of the values, where the weight assigned to each value is computed by a compatibility function of the query with the corresponding key \cite{vaswani2017attention}. The attention used in Transformer is scaled dot-product attention. The attention function is:
\begin{equation}
    Attention(Q,K,V) = softmax(\frac{QK^T}{\sqrt{d_k}})V
\end{equation}
where $Q$ is the query matrix, $K$ is the key matrix, $V$ is the value matrix.

Instead of performing a single attention calculation, the Transformer linearly projects the queries, keys and values $n$ times with different, learned linear projections, which is called multi-head attention. Given a query matrix $Q$, a key matrix $K$, and a value matrix $V$, the multi-head attention can be computed by
\begin{equation}
    MultiHead(Q,K,V)=Concat(h_1,h_2,\dots,h_n)W^O
\end{equation}
where $W^O$ is parameter matrix, $h_i$ ($i=1,\dots,n$) is attention head which can be computed by
\begin{equation}
    h_i = Attention(QW_i^Q,KW_i^K,VW_i^V)
\end{equation}
where $W_i^Q$, $W_i^K$, and $W_i^V$ are parameter matrices.

The fully connected feed-forward network in the encoder and decoder consists of two linear transformations with a ReLU activation \cite{nair2010rectified} in between. The linear transformations use different parameters from layer to layer. The feed-forward network can be expressed as
\begin{equation}
    FFN(x) = ReLU(xW_1+b_1)~W_2+b_2
\end{equation}
where $W_1$, $W_2$, $b_1$, and $b_2$ are parameters.

To use the sequence order information, the Transformer injects positional information by adding positional encoding to the input representations \cite{vaswani2017attention}. The positional encodings have the same dimension $d_{model}$ as the embeddings. The Transformer uses sine and cosine functions of different frequencies:
\begin{gather}
    PE_{(pos,2i)} = sin(pos/10000^{2i/d_{model}})\\
    PE_{(pos,2i+1)} = cos(pos/10000^{2i/d_{model}})
\end{gather}
where $pos$ is the position and $i$ is the dimension.

\subsection{Node correlation modeling}
\label{node_corr}
In this study, we use four graphs to describe the spatial adjacency, functional similarity, demographic similarity, and transportation similarity of nodes. The GCN model is based on these graphs. 

The spatial adjacency graph $G_{Adj}=(V,E_{Adj})$ is constructed by linking two geographically adjacent nodes $i$ and $j$ by an edge $e_{i,j} \in E_{Adj}$. Let $A_{Adj}$ be the adjacency matrix of graph $G_{Adj}$, and the element $a_{Adj}^{i,j} \in A_{Adj}$ is given by:
\begin{equation}
a_{Adj}^{i,j}=
    \begin{cases}
    1, ~\text{if node \textit{i} and node \textit{j} are adjacent}\\
    0, ~\text{otherwise.}
    \end{cases}
\end{equation}
\indent The functional similarity graph $G_{F}=(V,E_{F})$ is constructed by linking two nodes $i$ and $j$ that perform a similar function by an edge $e_{i,j} \in E_{F}$. The point of interest (POI) data is usually used to estimate the functional similarity between two nodes \cite{tang2021multi, geng2019spatiotemporal}. We calculate the POI similarity to quantify the functional similarity between two nodes and then determine whether the two nodes are functionally similar. The functional similarity graph is constructed using the same method in \cite{tang2021multi}. Let $A_{F}$ denote the adjacency matrix of graph $G_{F}$, the element $a_{F}^{i,j} \in A_{F}$ is given by:
\begin{equation}
a_{F}^{i,j}=
    \begin{cases}
    1, ~\text{if }sim(p_i,p_j)>d_F\\
    0, ~\text{otherwise.}
    \end{cases}
\end{equation}
\noindent where $p_i \in \mathbb{R}^{1 \times n}$ and $p_j\in \mathbb{R}^{1 \times n}$ are the vectors of POI count of nodes $i$ and $j$ respectively, $n$ is the number of POI categories, $sim(\cdot)$ is the calculation function of the Pearson coefficient, and $d_F$ is the threshold parameter, which is set to 0.8 \cite{tang2021multi}. 

The demographic similarity graph $G_{D}=(V,E_{D})$ is constructed by connecting two nodes $i$ and $j$ with similar demographic characteristics by an edge $e_{i,j} \in E_{D}$. The method we use to construct the demographic similarity graph is the same as the method to construct functional similarity graph. Let $A_{D}$ denote the adjacency matrix of graph $G_{D}$, the element $a_{D}^{i,j} \in A_{D}$ is given by:
\begin{equation}
a_{D}^{i,j}=
    \begin{cases}
    1, ~\text{if }sim(q_i,q_j)>d_D\\
    0, ~\text{otherwise.}
    \end{cases}
\end{equation}
where $q_i \in \mathbb{R}^{1 \times m}$ and $q_j\in \mathbb{R}^{1 \times m}$ are the demographic feature vectors of nodes $i$ and $j$ respectively, $m$ is the number of demographic features, $sim(\cdot)$ is the calculation function of the Pearson coefficient, and $d_D$ is the threshold parameter, which is set to 0.8.

Similarly, the transportation supply similarity graph $G_{T}=(V,E_{T})$ is constructed by connecting two nodes $i$ and $j$ with similar transportation supply characteristics by an edge $e_{i,j} \in E_{T}$. Let $A_{T}$ denote the adjacency matrix of graph $G_{T}$, the element $a_{T}^{i,j} \in A_{T}$ is given by:
\begin{equation}
a_{T}^{i,j}=
    \begin{cases}
    1, ~\text{if }sim(tr_i,tr_j)>d_T\\
    0, ~\text{otherwise.}
    \end{cases}
\end{equation}
where $tr_i \in \mathbb{R}^{1 \times m}$ and $tr_j\in \mathbb{R}^{1 \times m}$ are the demographic feature vectors of nodes $i$ and $j$ respectively, $m$ is the number of demographic features, $sim(\cdot)$ is the calculation function of the Pearson coefficient, and $d_T$ is the threshold parameter, which is set to 0.8.

\subsection{Spatio-Temporal Multi-Graph Transformer}

This section describes how the STMGT model operates based on the components mentioned in previous sections. Let $M$ denote the number of samples and $\textbf{X}_i$ denote the feature matrix (i.e., historical demand) of the $i$th sample. Let $k$ denote the number of Transformer blocks in the model. The operation process of STMGT model is presented in Algorithm~\ref{pseudo}. 

\begin{algorithm}[H]
	\caption{Algorithm of STMGT}\label{pseudo}
	\begin{algorithmic}[1]
	    \State \textbf{Input} Historical demand $\textbf{X}_t$,\\
	             \hspace{8.5mm} Spatial adjacency graph $G_{Adj}$,\\
	             \hspace{8.5mm} Functional similarity graph $G_F$,\\
	             \hspace{8.5mm} Demographic similarity graph $G_D$,\\
	             \hspace{8.5mm} Transportation supply similarity graph $G_T$,\\
	             \hspace{8.5mm} Weather conditoion $C$.
	   \State Concatenate the graphs: $G \leftarrow [G_{Adj},G_F,G_D,G_T]$
	   
        \State $\textbf{X}_{t,0}' \leftarrow \textbf{GCN}(\textbf{X}_t,G)$
        \State $\textbf{X}_{t,1}' \leftarrow \textbf{Transformer}(\textbf{X}_{t,0}',C)$
        \For{$i = 2,\dots,k$}
            \State $\textbf{X}_{t,i}' \leftarrow \textbf{Transformer}(\textbf{X}_{t,i-1}')$
        \EndFor
        \State $\widehat{\textbf{Y}}_t \leftarrow \textbf{Conv}(\textbf{X}_{t,k}')$
       
       \State \textbf{Output} prediction $\widehat{\textbf{Y}}_t$
	\end{algorithmic} 
\end{algorithm}

\section{Case Study}
\label{s:4}
In this section, we carried out case studies in Washington, D.C. and Austin, TX to evaluate the proposed STMGT model. The performance of STMGT was then compared with the state-of-the-art benchmark models.

\subsection{Data collection and description}
The data we used include the real-time dockless scooter-sharing trip OD data, the weather condition data, the POI count data, and the demographic and transportation supply data in the two case study sites. Table~\ref{tab:variable} presents the descriptive statistics of the input variables used in the case studies.

The real-time dockless scooter-sharing trip OD data was inferred from the General Bikeshare Feed Specification (GBFS) data \cite{nabsa2020} in Washington D.C. from June 19, 2019 to December 31, 2019. The data of four operators in Washington, D.C., including Bird, Lime, Lyft, and Spin, were collected. We collected the raw GBFS data using APIs provided by the vendors and then inferred the scooter trip origins and destinations using the algorithm developed by Xu et al.\cite{xu2020micromobility}. Note that we focused on trip generation (i.e., origin demand) prediction in this case study. We then aggregated the trip origins into the block group level and counted the trip origins at a 1-hour interval to obtain the hourly demand of each block group. The data from Jun 19, 2019 to Oct 22, 2019 was used as training set, the data from October 23, 2019 to November 22, 2019 was used as validation set, and the data from November 23, 2019 to December 31, 2019 was used as test set. The dockless scooter-sharing trip data in Austin, TX were collected from the data-sharing web portal \footnote{https://data.austintexas.gov/Transportation-and-Mobility/Shared-Micromobility-Vehicle-Trips/7d8e-dm7r} operated by the local government. The data from January 1, 2019 to April 30, 2019 was used as training set, the data from May 1, 2019 to May 31, 2019 was used as validation set, and the data from June 1, 2019 to June 30, 2019 was used as test set.


The weather condition data was collected from the \textit{Global Historical Climatology Network (GHCN)}\footnote{https://www.ncdc.noaa.gov/data-access/land-based-station-data/land-based-datasets/global-historical-climatology-network-ghcn} database. The daily precipitation, average temperature, and average wind speed were used in the STMGT model.

We collected the POI location data from the \textit{Open Data DC}\footnote{https://opendata.dc.gov/} portal, city of Austin open data portal \footnote{https://data.austintexas.gov/}, and the google map API \footnote{https://developers.google.com/maps}. The collected POIs included education facilities, recreational facilities, government facilities, medical facilities, automobile service facilities, financial service facilities, tourism attractions, hotels, and grocery stores. We counted the POIs of each category in each block group to aggregate the POIs into block-group-level.

The demographic data used in this study included the population density, the proportion of the young population, the proportion of the white population, the female proportion, the proportion of population with bachelor’s degree and above, the median household income, the proportion of households that own cars, and employment density. We also collected some transportation supply data, including bike lane density, WalkScore (an index to evaluate the quality of walking environment), transit stop density, parking lot density, and road network density. These variables are selected because of their high correlations to the dockless scooter-sharing demand \cite{bai2020dockless,merlin2021segment,noland2021scootin}. We collected these kinds of data from various sources. The demographic data was collected from the \textit{American Community Survey 2014-2018 5-year} estimates data. We used the \textit{Walkscore.com} API to obtain the WalkScore of a block group centroid and applied geographic information system (GIS) techniques to calculate bike lane density, transit stop density, parking lot density, and road network density.

\begin{table}[!t]
\caption{Descriptive statistics of input variables} 
\label{tab:variable}
\begin{adjustbox}{width=\columnwidth,center}
\begin{tabular}{lllllc}
\hline
\multirow{2}{*}{Variables}                                                                     & \multicolumn{2}{l}{Washington, D.C.} & \multicolumn{2}{l}{Austin, TX} & \multirow{2}{*}{Graph}                                                                                   \\
                                                                                               & Mean              & Std.             & Mean           & Std.          &                                                                                                          \\ \hline
No. of education facilities                                                                 & 0.34              & 0.66             & 0.66           & 1.06          & \multirow{9}{*}{\begin{tabular}[c]{@{}c@{}}Functional \\ similarity \\ graph\end{tabular}}               \\
No. of recreational facilities                                                              & 0.33              & 0.75             & 0.16           & 0.42          &                                                                                                          \\
No. of government facilities                                                                & 10.93             & 30.83            & 5.28           & 22.58         &                                                                                                          \\
No. of medical facilities                                                                   & 0.41              & 1.04             & 0.78           & 1.59          &                                                                                                          \\
No. of auto service facilities                                                        & 0.09              & 0.36             & 0.88           & 1.68          &                                                                                                          \\
No. of financial service facilities                                                         & 0.47              & 2.18             & 1.30           & 3.14          &                                                                                                          \\
No. of tourism attractions                                                                  & 0.07              & 0.96             & 0.22           & 0.78          &                                                                                                          \\
No. of hotels                                                                               & 0.38              & 1.28             & 0.32           & 1.31          &                                                                                                          \\
No. of grocery stores                                                                       & 0.18              & 0.45             & 0.59           & 0.91          &                                                                                                          \\ \hline
Population density (per sq. mile)                                                              & 21,029            & 16,424           & 4,798          & 3,937         & \multirow{8}{*}{\begin{tabular}[c]{@{}c@{}}Demographic \\ similarity \\ graph\end{tabular}}              \\
Pct. of the young population                                                             & 32\%              & 16\%             & 45\%           & 12\%          &                                                                                                          \\
Pct. of the white population                                                             & 41\%              & 33\%             & 74\%           & 15\%          &                                                                                                          \\
Female proportion                                                                              & 53\%              & 7\%              & 49\%           & 6\%           &                                                                                                          \\
\begin{tabular}[c]{@{}l@{}}Pct. of population with BA's \\ degree and above\end{tabular} & 56\%              & 30\%             & 51\%           & 22\%          &                                                                                                          \\
Median household income (USD)                                                            & 96,519            & 55,202           & 80,644         & 37,516        &                                                                                                          \\
Pct. of households own cars                                                              & 68\%              & 19\%             & 91\%           & 8\%           &                                                                                                          \\
Employment density (per mi$^2$)                                                              & 12,230            & 11,817           & 2,748          & 2,043         &                                                                                                          \\ \hline
Bike lane density (mi / mi$^2$)                                                          & 11.47             & 13.34            & 6.37           & 6.46          & \multirow{5}{*}{\begin{tabular}[c]{@{}c@{}}Transportation \\ supply \\ similarity \\ graph\end{tabular}} \\
WalkScore                                                                                      & 73.47             & 21.84            & 33.55          & 27.82         &                                                                                                          \\
Transit stop density (per mi$^2$)                                                            & 571.08            & 408.86           & 12.77          & 14.41         &                                                                                                          \\
Parking lot density (per mi$^2$)                                                             & 142.48            & 106.43           & 0.86           & 11.72         &                                                                                                          \\
Road network density (mi / mi$^2$)                                                       & 53.24             & 20.28            & 14.44          & 12.20         &                                                                                                          \\ \hline
Daily precipitation (inch)                                                                     & 0.11              & 0.35             & 0.13           & 0.46          & \multirow{3}{*}{-}                                                                                       \\
Average temperature (°F)                                                                       & 65.76             & 16.55            & 66.18          & 13.50         &                                                                                                          \\
Average wind speed (m/s)                                                                       & 7.88              & 2.75             & 6.44           & 2.71          &                                                                                                          \\ \hline
\end{tabular}
\end{adjustbox}
\end{table}

\subsection{Model setting}
The case studies were conducted using an NVIDIA 1080Ti GPU. After hyperparameter tuning, the batch size was set to 36. The number of Transformer blocks was set to 3. The input time sequence length was 24. The model was trained with $L2$ loss using the Adam optimizer \cite{kingma2014adam} for 300 epochs. The initial learning rate was set to 0.005.

\subsection{Models comparison}
In this section, we compared the proposed STMGT model with several benchmark models. The details of these models are described as follows. Note that all the models were fine-tuned.

\begin{itemize}
    \item \textbf{HA}: Historical Average is one of the most fundamental statistical models for time series prediction. HA predicts the demand in a specific time period by averaging historical observations. 
    \item \textbf{ARIMA}: Auto-Regressive Integrated Moving Average is a statistical time series prediction model. ARIMA fits a parametric model based on historical observations to predict future demand. The order of ARIMA was set to (1,0,0) in the case study.
    \item \textbf{SVR}: Support Vector Regression \cite{smola2004tutorial} is a machine learning model that uses the same principle as Support Vector Regression (SVM) but for regression problems. We used the Radial Basis Function kernel here. The cost was set to 1, and the gamma was set to 0.02.
    \item \textbf{GBDT}: Gradient Boosting Decision Tree \cite{friedman2001greedy} is a tree-based ensemble machine learning model. In this case study, the number of trees was set to 2000, the maximum depth was set to 7, and the learning rate was set to 0.05. 
    \item \textbf{RF}: Random Forest \cite{breiman2001random} is another tree-based ensemble machine learning method. In this model, the number of trees was set to 110, and the number of features to consider when looking for the best split was set to 7.
    \item \textbf{MLP}: Multiple Layer Perceptron is a classical feedforward artificial neural network. In the case study, we used an MLP model with 100 neurons in the hidden layer. The activation function was ReLU. The learning rate was set to 0.001.  
    \item \textbf{GRU}: Gated Recurrent Unit \cite{cho2014properties} is a widely used RNN model for time series modeling. The learning rate was set to 0.001, the batch size was set to 64, and the number of hidden units was 32.
    \item \textbf{LSTM}: Long Short-Term Memory \cite{hochreiter1997long} is another widely used neural network based on the gating mechanism. In this model, the learning rate was 0.001, the batch size was 64, and the number of hidden units was 32. 
    \item \textbf{Transformer}: Transformer is a attention-based deep learning network\cite{vaswani2017attention}. In this model, the learning rate was set to 0.005, the batch size was 36, and the number of blocks was 3.  
    \item \textbf{Informer}: Informer \cite{zhou2021informer} is a transformer-based model for long sequence time-series forecasting. With a ProbSparse self-attention mechanism, Informer achieves higher computation efficiency and comparable performance. In this model, the learning rate was set to 0.0001, the batch size was 32, the number of heads is 8.
    \item \textbf{T-GCN}: Temporal Graph Convolutional Network \cite{zhao2019t} is a spatiotemporal graph convolutional neural network that captures spatial and temporal dependency simultaneously. The spatial adjacency graph $G_{Adj}$ was used as the input of the GCN model in the T-GCN model. The learning rate was set to 0.001. The batch size was set to 64. The number of hidden units was 32.
    \item \textbf{STGCN}: Spatio-Temporal Graph Convolutional Networks \cite{yu2018spatio} formulates the forecasting problem on graphs and uses complete convolutional structures, which enable much faster training speed with fewer parameters. The learning rate was set to 0.001. The batch size was 64.
    \item \textbf{AGCRN}: Adaptive Graph Convolutional Recurrent Network \cite{bai2020adaptive} is a deep learning network that captures fine-grained spatial and temporal correlations in traffic series automatically based on recurrent networks and two adaptive modules, named Node Adaptive Parameter Learning (NAPL) module (to capture node-specific patterns) and Data Adaptive Graph Generation (DAGG) module (to infer the inter-dependencies). In this model, the learning rate was set to 0.001. The batch size was 64. The number of hidden units was 64.
    \item \textbf{MC\_STGCN}: The Multi-Community Spatio-Temporal Graph Convolutional Network \cite{tang2021multi} is an advanced spatiotemporal graph convolutional neural network for passenger demand prediction. The model was trained using Adam optimizer with learning rate of 0.001. The batch size was set to 64. The number of hidden units was 32.
   
\end{itemize}

We evaluated the performance of the models using Mean Absolute Error (MAE), Root Mean Square Error (RMSE), Mean Absolute Percentage Error (MAPE), and Symmetric Mean Absolute Percentage Error (SMAPE) \cite{flores1986pragmatic}. Note that MAPE can be greatly affected by the small values. We only calculate MAPE for samples with demand no less than 10, named MAPE$_{10}$ \cite{tang2021multi}.



\begin{table*}[!t]
\caption{Performance of the STMGT model and the benchmark models} 
\label{tab:performance}
\begin{center}
\begin{adjustbox}{width=1.8\columnwidth,center}
\begin{tabular}{l|llll|llll}
\hline
\multirow{2}{*}{Methods} & \multicolumn{4}{c|}{Washington, D.C.}                                          & \multicolumn{4}{c}{Austin, TX}                                                \\
                         & \multicolumn{1}{c}{MAE} & \multicolumn{1}{c}{RMSE} & \multicolumn{1}{c}{MAPE$_{10}$}  & \multicolumn{1}{c|}{SMAPE} & \multicolumn{1}{c}{MAE} & \multicolumn{1}{c}{RMSE} & \multicolumn{1}{c}{MAPE$_{10}$} & \multicolumn{1}{c}{SMAPE} \\ \hline
HA                       & 0.3675                  & 0.7031                   & 0.5678             & 0.5620                        & 0.1942                   & 0.7420                        & 0.6563            & 0.3120                \\
ARIMA                    & 0.5591                  & 0.9999                   & 0.8530          & 0.7750           & 0.3913                   & 0.9997                        & 0.7879            & 0.5537                        \\
SVR                      & 0.2906                  & 0.6993                   & 0.4286           & 0.3083                       & 0.1959                   & 0.7248                        & 0.3580           & 0.2877                        \\
GBDT                     & 0.3579                  & 0.6832                   & 0.3466          & 0.2571                       & 0.1290                       & 0.7126                        & 0.3814           & 0.2196                       \\
RF                       & 0.3714                  & 0.7160                   & 0.3418          & 0.2603                       & 0.1888                   & 0.7185                        & 0.3316           & 0.2238                        \\
MLP                      & 0.3542                  & 0.6867                   & 0.4001             & 0.2597                       & 0.2011                   & 0.7190                        & 0.3972           & 0.2573                       \\
GRU                      & 0.2079                  & 0.5286                   & 0.3388            &  0.2398                      & 0.0954                       & 0.4313                        & 0.3332           &  0.2026                      \\
LSTM                     & 0.2269                  & 0.5403                   & 0.3401             & 0.2455                       & 0.1024                       & 0.4541                       & 0.3351            & 0.2056                        \\
Transformer         & 0.1809                  & 0.4935                   & 0.3332              & 0.2319                       & 0.0899                       & 0.4145                       & 0.3310           & 0.1998                       \\
Informer                 & 0.1392                  & 0.3186                   & 0.3189           & 0.2301                       & 0.0639                       & 0.3212                        & 0.3049           &  0.1835                      \\
T-GCN                    & 0.1308                  & 0.3543                   & 0.3222            & 0.2235                       & 0.0756                       & 0.3618                        & 0.3190            &  0.1901                      \\
STGCN                    & 0.1192                  & 0.3241                   & 0.3100           & 0.2156                       & 0.0683                       & 0.3582                        & 0.3012          & 0.1897                       \\
AGCRN                    & 0.0881                  & 0.2866                   & 0.2950           & 0.1889                       & 0.0515                       & 0.2438                        & 0.2690          & 0.1669   \\
MC\_STGCN                & 0.1121                  & 0.3492                   & 0.3198          & 0.2103                       & 0.0631                       & 0.3592                        & 0.3001           & 0.1810                  \\
\textbf{STMGT}                    & \textbf{0.0614}                  & \textbf{0.2215}                   & \textbf{0.2690}             & \textbf{0.1711}                       & \textbf{0.0363}                       & \textbf{0.1930}                        & \textbf{0.2416}           & \textbf{0.1590}                      \\ \hline
\end{tabular}
\end{adjustbox}
\end{center}
\end{table*}

The performance of these models is shown in Table~\ref{tab:performance}. We can see that the STMGT model outperformed all the benchmark models. The RNN models (i.e., GRU and LSTM) had better prediction performance than the statistical models (i.e., HA and ARIMA), the classical machine learning models (i.e., SVR, GBDT, and RF), and the classical neural network model (i.e., MLP). Among the two RNN models, GRU slightly outperformed LSTM. The attention-based models (Transformer, Informer) outperformed the RNN models. The spatiotemporal neural network models (i.e., T-GCN, MC\_STGCN, STGCN, AGCRN, STMGT) significantly outperformed the RNN models, which can only capture temporal dependency. This result indicated that there exist strong spatial correlations among nodes, and the GCN components with well-designed graphs can well capture these correlations. The multi-graph model (i.e., MC\_STGCN and STMGT) had better performance than T-GCN, which took single-graph input. This implied that the more comprehensive spatial information provided by multiple graphs with a well-tuned fusing scheme could improve the model performance. The performance of STMGT model was better than MC\_STGCN, which used GRU to capture temporal dependency and did not take weather conditions into account. This result indicated that using Transformer instead of RNN-based model and considering key factors that are correlated to demand can improve the demand forecasting accuracy.

\subsection{Ablation study}

We conducted an ablation study for the proposed model. The ablation study examines the performance of the model by removing certain components to see the contribution of the removed components \cite{meyes2019ablation}. In this ablation study, we generated five models by removing spatial adjacency graph, functional similarity graph, demographic similarity graph, transportation supply similarity graph, or weather information, respectively. 

The performance of the five ablated models is presented in Table~\ref{tab:ablation}. The results suggested that the weather information, the spatial adjacency graph, the functional similarity graph, and the demographic and built environment contributed to the prediction accuracy, and the weather information contributed the most. According to the results, the prediction error increased a lot when removing the weather information (MAE increased 57\% and RMSE increased 40\% for Washington D.C., MAE increased 26\% and RMSE increased 34\% for Austin, TX). This result indicates that the weather information can greatly improve the model performance. It is reasonable because the use of dockless scooter-sharing is sensitive to the weather condition. For example, the use of dockless scooter-sharing will significantly decrease when it is cold and rainy \cite{noland2021scootin}. Additionally, a relatively small increase in prediction error (20\% for MAE and 6\% for RMSE in Washington, D.C., 11\% for MAE and 4\% for RMSE in Austin, TX) occurred when we removed the spatial adjacency graph. This indicates that although the spatial adjacency graph improved the model performance, the contribution of this component was relatively limited. The prediction accuracy of the model also decreased when we removed the functional similarity graph (23\% for MAE and 27\% for RMSE in Washington, D.C., 24\% for MAE and 8\% for RMSE in Austin, TX), the demographic similarity graph (44\% for MAE and 20\% for RMSE in Washington, D.C., 20\% for MAE and 8\% for RMSE in Austin, TX), and the transportation supply similarity graph (44\% for MAE and 26\% for RMSE in Washington, D.C., 26\% for MAE and 8\% for RMSE in Austin, TX), which suggests that these components can enhance the model performance. 

\begin{table}[H]
\caption{Results of ablation study} 
\label{tab:ablation}
\begin{center}
\begin{adjustbox}{width=\columnwidth,center}
\begin{tabular}{lllllll}
\hline
\multirow{2}{*}{Methods}                                                        & \multicolumn{3}{c}{Washington, D.C.}                                          & \multicolumn{3}{c}{Austin, TX}                                                \\
                                                                                & \multicolumn{1}{c}{MAE} & \multicolumn{1}{c}{RMSE} & \multicolumn{1}{c}{MAPE$_{10}$} & \multicolumn{1}{c}{MAE} & \multicolumn{1}{c}{RMSE} & \multicolumn{1}{c}{MAPE$_{10}$} \\ \hline
STMGT                                                                           & 0.0614                       & 0.2215                        & 0.2690                        & 0.0363                       & 0.1930                        & 0.2416                        \\
\begin{tabular}[c]{@{}l@{}}w/o Spatial \\ adjacency\end{tabular}                & 0.0739                       & 0.2352                        & 0.3035                        & 0.0404                       & 0.2008                        &  0.2585                       \\
\begin{tabular}[c]{@{}l@{}}w/o Functional \\ similarity\end{tabular}            & 0.0757                       & 0.2822                        & 0.3036                        & 0.0450                       & 0.2093                        & 0.2689                        \\
\begin{tabular}[c]{@{}l@{}}w/o Demographic \\ similarity\end{tabular}           & 0.0886                       & 0.2654                        & 0.3101                        & 0.0434                       & 0.2076                        & 0.2688                        \\
\begin{tabular}[c]{@{}l@{}}w/o Transportation \\ supply similarity\end{tabular} & 0.0882                       & 0.2780                        & 0.3033                        & 0.0457                       & 0.2090                        & 0.2600                        \\
\begin{tabular}[c]{@{}l@{}}w/o Weather \\ information\end{tabular}              & 0.0967                       & 0.3097                        & 0.3168                        & 0.0459                       & 0.2592                        & 0.2956                        \\ \hline
\end{tabular}
\end{adjustbox}
\end{center}
\end{table}

\subsection{Permutation Feature Importance}
\begin{figure*}[hbt!]
    \centering
    \includegraphics[width=0.9\textwidth]{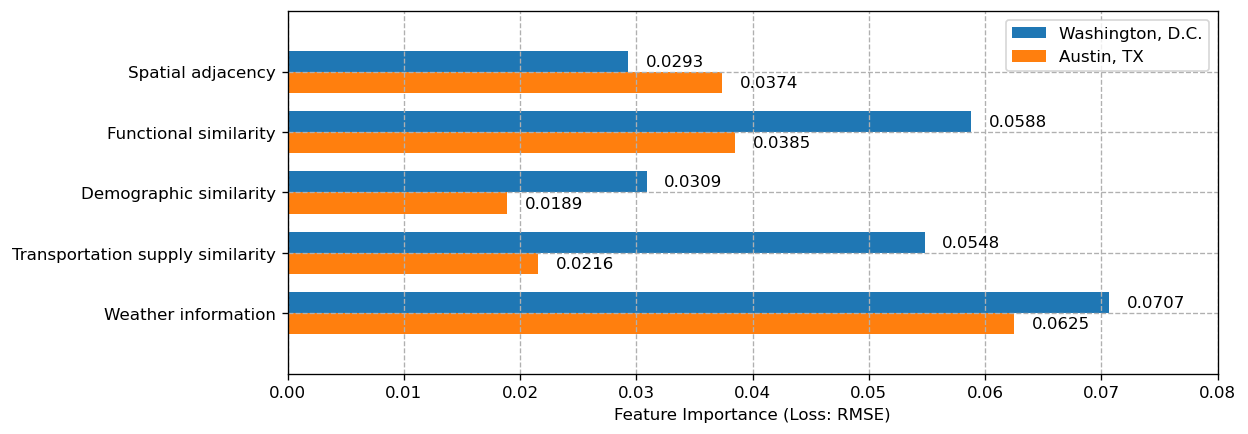}
    \caption{Permutation Feature Importance}
    \label{fig:fi}
\end{figure*}

We used the permutation feature importance \cite{fisher2019all} to further explore the impacts of different model components on the overall performance. Permutation feature importance measures the importance of a feature by calculating the increase of error after permuting the feature. The permutation feature importance of each model component is presented in Figure~\ref{fig:fi}. According to the results, the most important feature is weather information (RMSE increases by 0.0707 for Washington, D.C. dataset and 0.0625 for Austin, TX dataset after permutation), which is consistent with the results of ablation study. The second important feature is the functional similarity (RMSE increases by 0.0588 for Washington, D.C. dataset and 0.0385 for Austin, TX dataset after permutation). The impact of the remaining three features (i.e., spatial adjacency, demographic similarity, and transportation supply similarity) is different in the two case studies. Spatial adjacency is more important in the Austin, TX case, while demographic similarity and transportation supply similarity are more important in the Washington, D.C. case. The RMSE increased by at least 0.0189 after the permutation, which indicates that all features contributed to the prediction accuracy significantly. This result is also consistent with the ablation study.

\subsection{Prediction results}
\begin{figure*}[!t]
\centering
\subfloat[]{\includegraphics[width=0.485\textwidth]{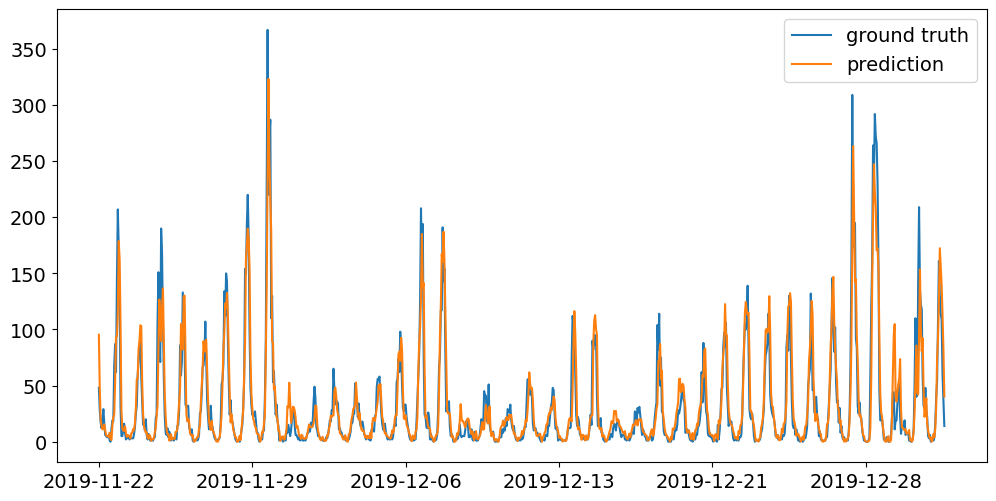}}%
\label{fig:region1}
\subfloat[]{\includegraphics[width=0.5\textwidth]{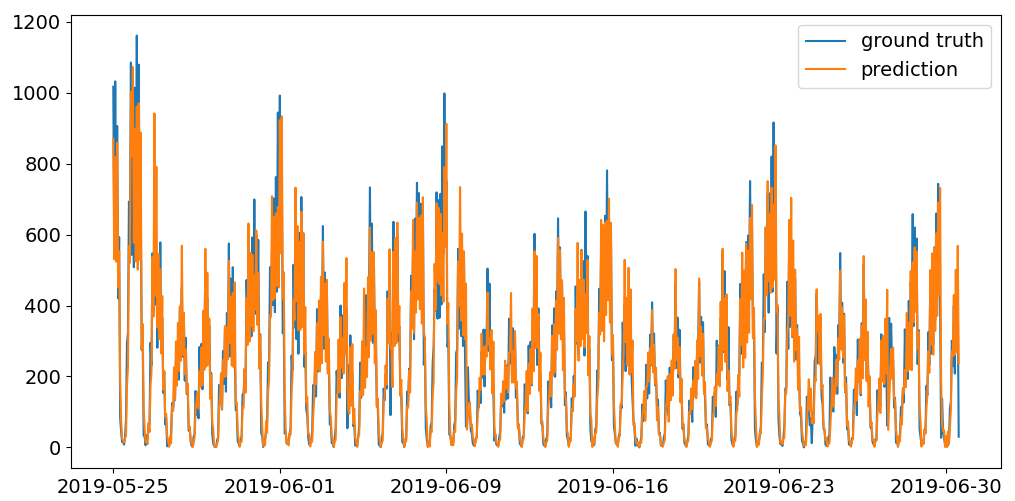}}%
\label{fig:region3}
\caption{Comparison of STMGT model prediction and ground truth demand. (a) Region 1, Washington, D.C. (b) Region 2, Austin, TX}
\label{fig:result}
\end{figure*}

We further compared the prediction of the STMGT model and the observed demand using test data for two regions (one in Washington, DC, the other in Austin, TX) with high variance and different demand volumes. The results are presented in Figure~\ref{fig:result}. The figures show the temporal fluctuation of the ground truth and predicted hourly dockless scooter-sharing demand in two different areas. As we can see, the temporal distributions of the dockless scooter-sharing demand in the two regions had different demand intensities. The temporal distributions of the observed demand and the model prediction are largely similar. Although the model sometimes cannot fully capture extreme values, the model prediction can effectively follow the temporal fluctuation of observed data in different regions and different time periods. It is worth noting that there were obvious peaks of demand in these figures. These peaks usually occurred over the weekends, and the proposed model performed well in predicting these peak demands. 

\begin{figure*}[!t]
\centering
\subfloat[]{\includegraphics[width=0.49\textwidth]{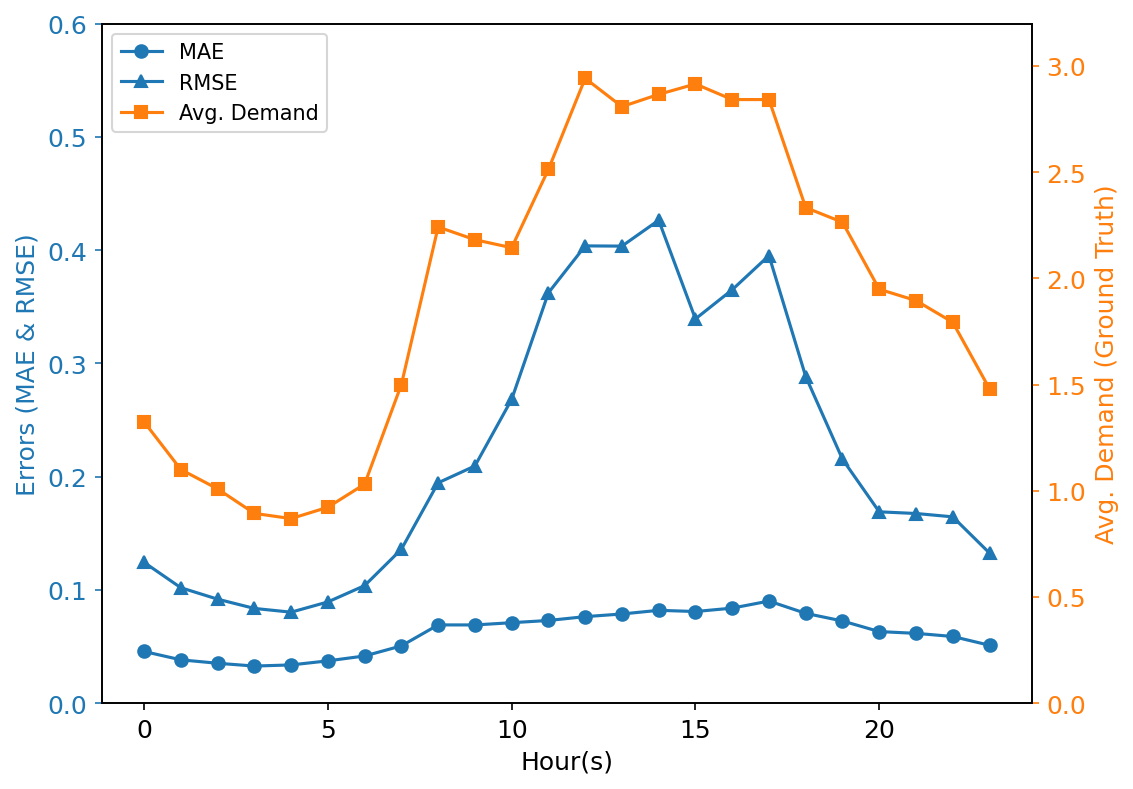}}%
\label{fig:dc_err}
\subfloat[]{\includegraphics[width=0.49\textwidth]{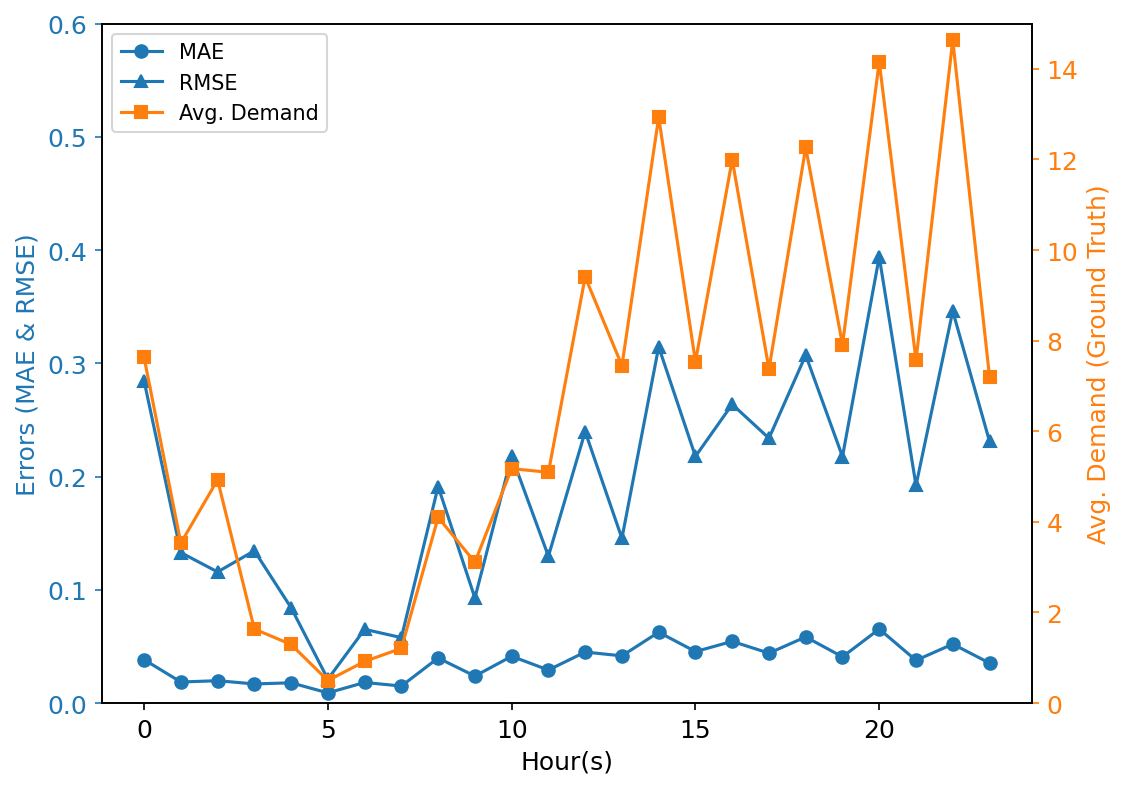}}%
\label{fig:austin_err}
\caption{Average prediction errors at different hours of the day. (a) Washington, D.C. (b) Austin, TX}
\label{fig:error}
\end{figure*}

We further examined the average prediction errors (i.e., MAE and RMSE) for different hours of the day. The results are presented in Figure~\ref{fig:error}. Note that we also present ground-truth average demand for the zones (i.e., census block groups or census tracts) in each hour. According to Figure~\ref{fig:error}, the prediction errors increase as the ground-truth average demand increases. For the Washington, D.C. dataset, the MAE is always smaller than 0.1 and the RMSE is always below 0.45. For the Austin, TX dataset, the MAE is always smaller than 0.07 and the RMSE is always below 0.40.

\section{Discussion and Conclusion}
\label{s:5}
In this study, we proposed a novel deep learning architecture \textit{Spatio-Temporal Multi-Graph Transformer} (STMGT) for real-time dockless scooter-sharing demand forecasting. The proposed STMGT incorporated zonal context and weather conditions for spatiotemporal demand prediction. It used a GCN component to capture spatial dependency and a Transformer component to capture temporal dependency. The proposed model outperformed all selected benchmark models in the two real-world case studies. The outstanding performance of STMGT was due to following reasons. First, the Transformer component was effective in capturing long-range temporal dependency from sequential data, thus outperformed the RNN-based spatiotemporal models. Second, the graph-based deep learning model incorporated comprehensive spatial dependency based on the graphs that were constructed using transportation domain knowledge. Third, weather conditions, which could greatly influence the dockless scooter-sharing demand, were included in the model.


Particularly, the graph-based deep learning models are effective in capturing spatial dependency. STMGT utilized a graph-based deep learning model (i.e., GCN) to capture spatial dependency. We constructed four graphs (i.e., spatial adjacency graph, functional similarity graph, demographic similarity graph, and transportation supply graph) to represent the spatial correlations between areas. The GCN operated on these graphs and attached the spatial dependency to the time sequence data (i.e., historical demand). 


To explore the impacts of different features, we conducted an ablation study and calculated the permutation feature importance, and found that both approaches led to consistent findings. More specifically, the results showed that the most important model component was the weather information, which was not considered by most existing travel demand prediction studies. The prediction accuracy of the model decreased when we ablate the weather information or one of the graphs. The RMSE increased by at least 0.0189 after we permuted the weather information or one of the graphs. The results of ablation study and permutation feature importance  indicated that the weather information and the graphs contributed to the prediction accuracy significantly.  

Although the proposed STMGT model has achieved high prediction accuracy, several limitations still require future work. First, we assumed that the spatial dependency is fixed in this study. However, in the real-world scenarios, spatial dependency can be dynamic with different time steps. For example, the frequency of transit during peak hours and night are significantly different. We may use dynamic graphs to capture the dynamical spatial dependency in future work.
Second, we used daily weather condition data to forecast the hourly dockless scooter-sharing demand in this study. Weather condition data with a smaller time interval (e.g., hourly) may be used to reflect more precise weather conditions in the STMGT and potentially achieve a better predictive performance in future work.

\section*{Acknowledgments}
This research was supported by the U.S. Department of Transportation through the Southeastern Transportation Research, Innovation, Development and Education (STRIDE) Region 4 University Transportation Center (Grant No. 69A3551747104) and by the University of Florida AI Research Catalyst Fund.



\bibliographystyle{IEEEtran}
\bibliography{sample.bib}

 



\section*{Biography}
\vspace{-30pt}
\begin{IEEEbiography}[{\includegraphics[width=1in,height=1.25in,clip,keepaspectratio]{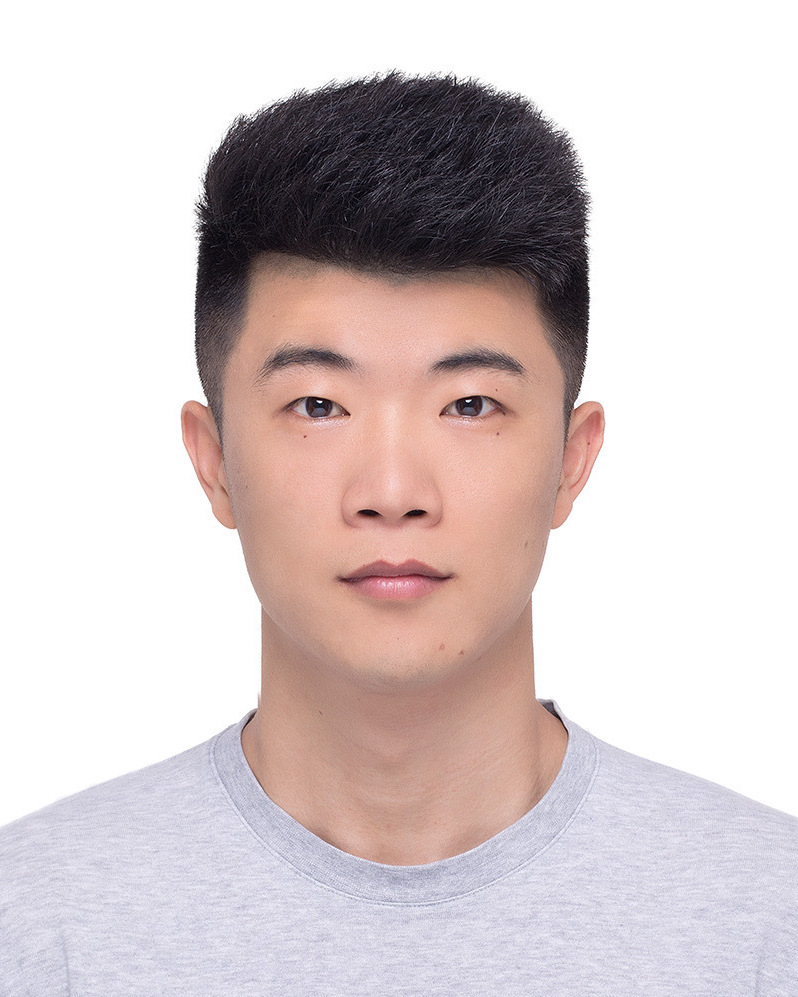}}]{Yiming Xu}
 is a Ph.D. candidate in the Department of Civil and Coastal Engineering at the University of Florida. He received his B.E. and M.E. degrees in transportation engineering from Tongji University, China. His research interests include shared mobility, machine learning, deep learning, and data analytics for transportation systems.
\end{IEEEbiography}


\begin{IEEEbiography}[{\includegraphics[width=1in,height=1.25in,clip,keepaspectratio]{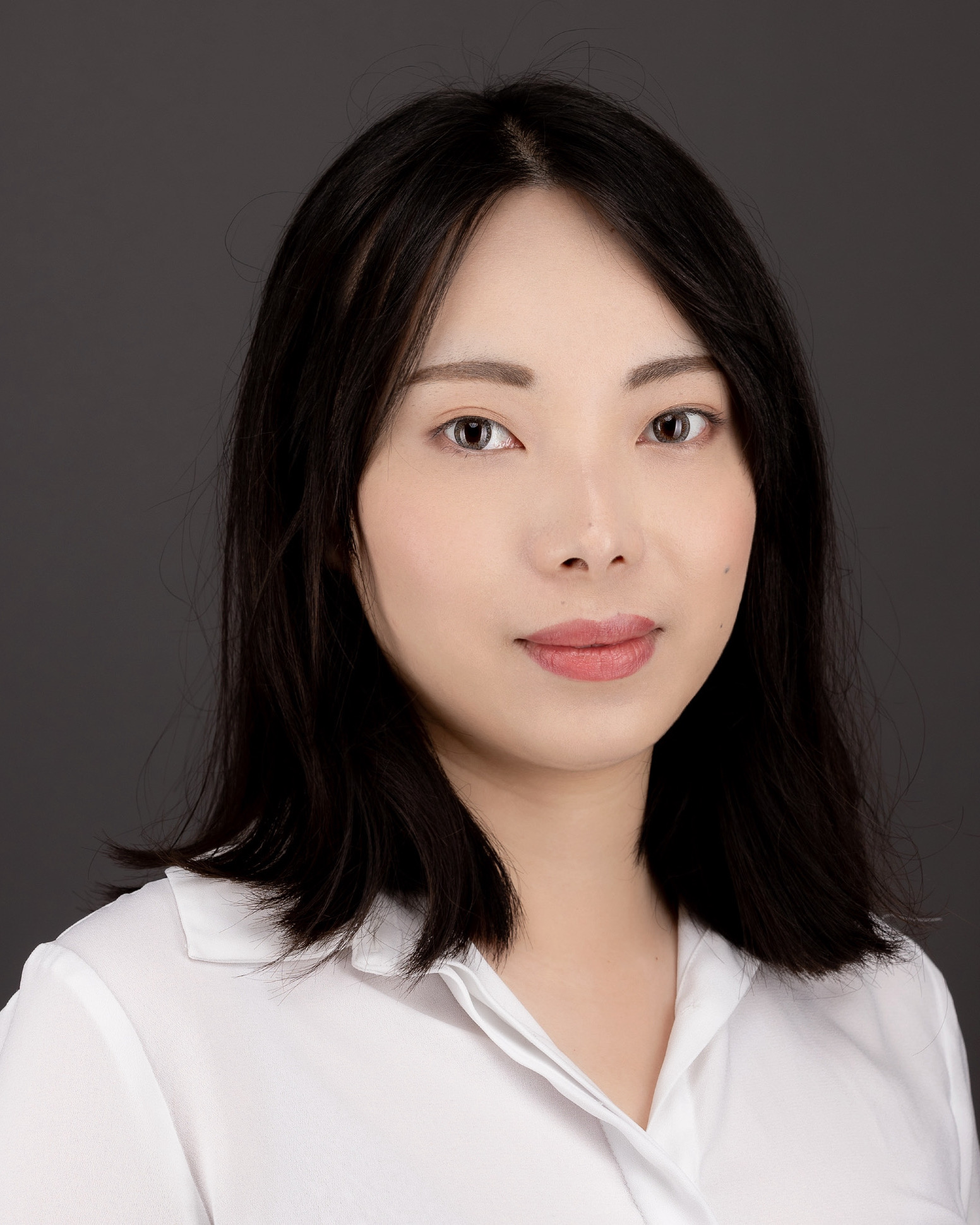}}]{Xilei Zhao} received a B.E. degree in civil engineering from Southeast University, Nanjing, China, in 2013. She then received a M.S.E. degree in civil engineering in 2016 and a M.S.E. degree in applied mathematics and statistics and a Ph.D. degree in civil engineering in 2017 from the Johns Hopkins University, Maryland, USA. She is currently an Assistant Professor in the Department of Civil and Coastal Engineering at the University of Florida, where she leads the Smart, Equitable, Resilient Mobility Systems (SERMOS) Lab. Prior to her appointment at the University of Florida, she worked as a Postdoctoral Fellow in the School of Industrial and Systems Engineering at Georgia Tech (2018--2019) and a Research Fellow in the Department of Industrial and Operations Engineering at the University of Michigan (2017--2018). Zhao’s work focuses on developing and applying data and computational science methods to tackle challenging problems in transportation and resilience. She specializes in big data analytics and trustworthy AI applications in travel behavior modeling, modeling and planning evacuation, and quantifying resilience for critical infrastructure systems, societal systems, and communities.
\end{IEEEbiography}


\begin{IEEEbiography}[{\includegraphics[width=1in,height=1.25in,clip,keepaspectratio]{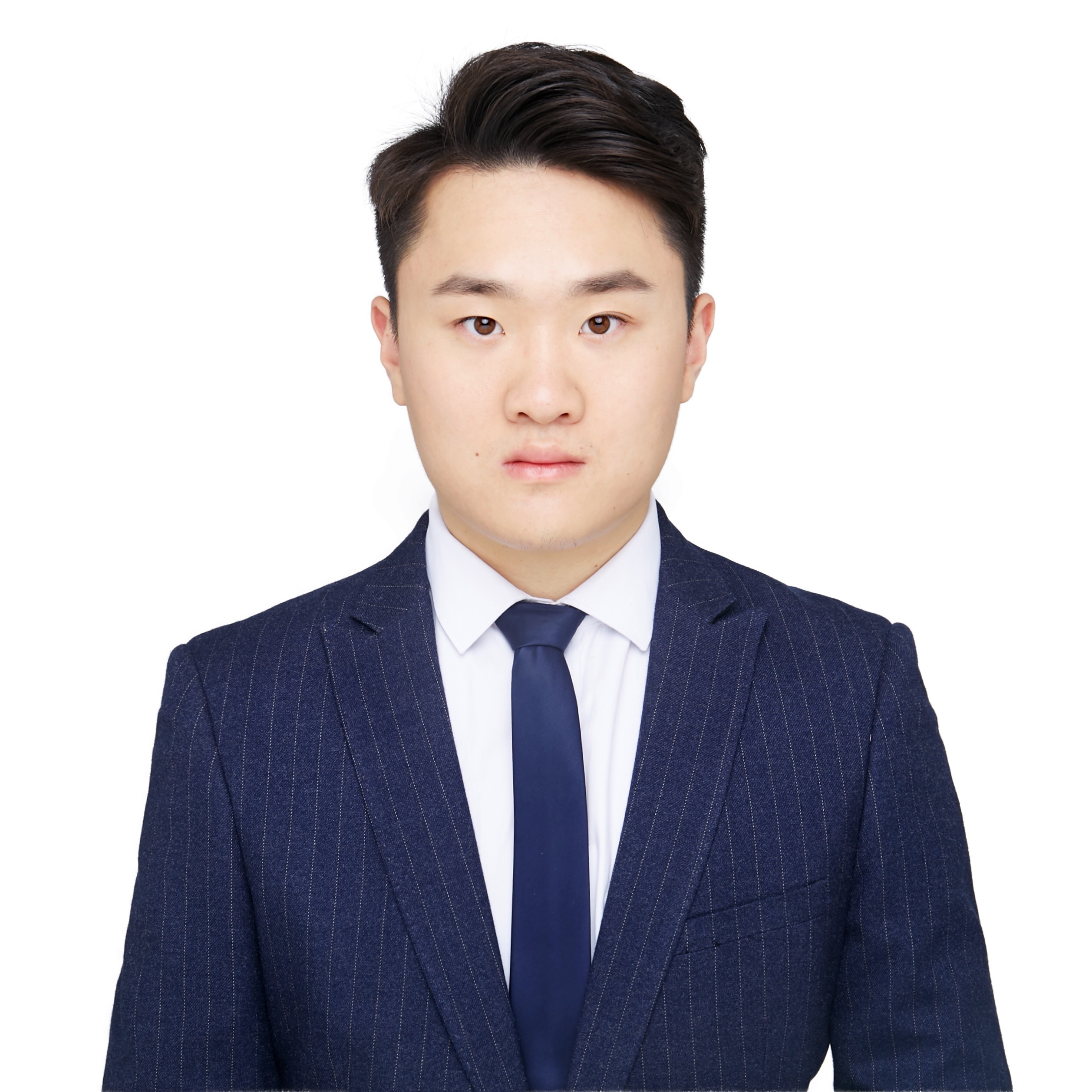}}]{Xiaojian Zhang}
is a Ph.D. student in the Department of Civil and Coastal Engineering at the University of Florida. He received his B.E. degree in Southwest Jiaotong University, China. He is currently working on promoting fairness in artificial intelligence and applying machine learning methods for earthquake evacuation decision makings.
\end{IEEEbiography}

\begin{IEEEbiographynophoto}{Mudit Paliwal}
 is a senior product management analyst at WR Berkley. As an experienced Operations engineer in commercial shipping, Mudit has worked across multiple international disciplines in both oil and gas transportation with giants like Maersk, Total and NYK. In 2021, Mudit graduated with a Masters in Industrial and Systems Engineering from the University of Florida, concentrating in Data Analytics and Systems Thinking. During his Masters, he developed his interests in deep learning and transfer learning which helped him leverage his research on forecasting models for micromobility and  behavior analysis of commercial aviation pilots for a project sponsored by NASA Langley Research Center.
\end{IEEEbiographynophoto}

\vfill

\end{document}